\documentclass[fleqn,usenatbib]{mnras}

\usepackage{newtxtext,newtxmath}

\usepackage[T1]{fontenc}

\DeclareRobustCommand{\VAN}[3]{#2}
\let\VANthebibliography\thebibliography
\def\thebibliography{\DeclareRobustCommand{\VAN}[3]{##3}\VANthebibliography}

\usepackage{graphicx}	
\usepackage{amsmath}	

\title[Drunk fireworks in Fallas]{The popular myth of the drunk fireworks in the Valencian Fallas: “If you run, it chases you”}

\author[Eloy Peña-Asensio]{
Eloy Peña-Asensio$^{1,2}$\thanks{E-mail: eloy.pena@uab.cat}
\\
$^{1}$Universitat Autònoma de Barcelona 
08193 Bellaterra, Catalonia, Spain\\
$^{2}$Institut de Ciències de l’Espai (ICE, CSIC), C/ de Can Magrans s/n, 
08193 Cerdanyola del Vallès, Catalonia, Spain\\
}

\date{Accepted April 1, 2022.}

\pubyear{2022}

\begin{document}
\label{firstpage}
\pagerange{\pageref{firstpage}--\pageref{lastpage}}
\maketitle

\begin{abstract}

There is a popular myth transferred from generation to generation related to the festivity of the Valencian Fallas. During these days, it is traditional to throw pyrotechnic devices of all kinds. There is a legend about the so-called drunk fireworks, small rockets that move randomly at high speed. According to folk wisdom, if you try to run away from a drunk firework, it will chase you until it hits you. It is performed a Monte Carlo simulation without considering aerodynamic effects to see if there is indeed a greater chance of a person being hit when moving. The results indicate that the popular myth is correct, at least in terms of having a higher probability of impact. Once again, it is demonstrated that popular myths can be a source of valuable knowledge.

\end{abstract}

\begin{keywords}
Fallas -- Festivity -- Impacts -- Drunk fireworks -- Monte Carlo
\end{keywords}



\section{What are the Valencian Fallas?}

\begin{figure}
	\includegraphics[width=\columnwidth]{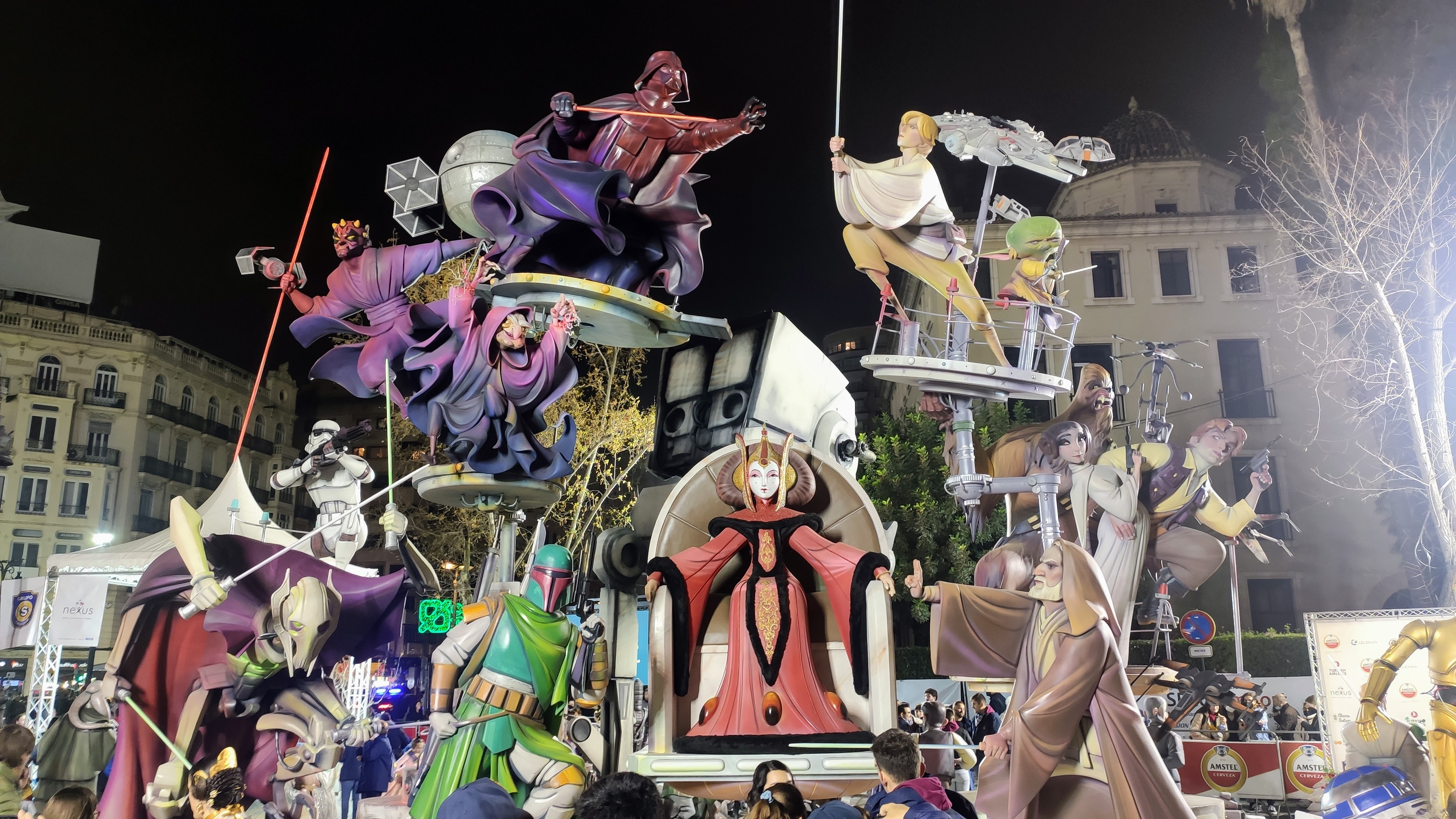}
	\centering
	\caption{Falla inspired by the Star Wars universe. Valencia north train station, 2022.}
	\label{fig:figFallas1}
\end{figure}

\begin{figure}
	\includegraphics[width=\columnwidth]{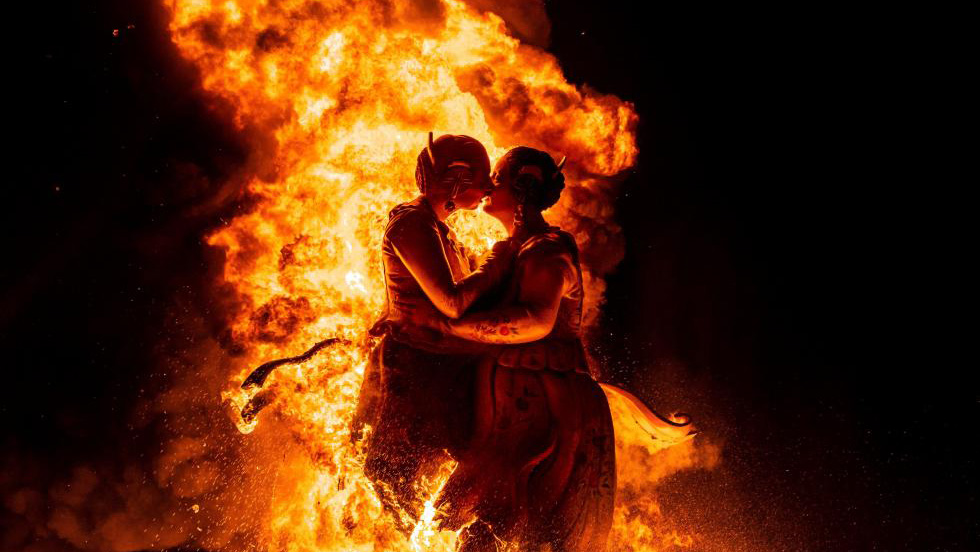}
	\centering
	\caption{Falla supporting the right to sexual freedom by visualizing lesbian relationships with two falleras in traditional dress \citep{lopez2020vestido}. 2021.}
	\label{fig:figFallas2}
\end{figure}

Fallas are the most important festivities of the city of Valencia in honour of the patron saint Sant Josep and are also celebrated in many towns of the Valencian Country as Alzira or Torrent \citep{costa2001festivity}. They are held between March 15, the day of the so-called "plantà" of the falleros's monuments, and March 19, on whose night the "cremà" is celebrated. The falla is typically composed of a wooden structure that holds cork figures with different themes, see Figure \ref{fig:figFallas1}. They are usually satirical in nature about current affairs \citep{rius2021traditional}. The last day of the festivity and after choosing the best ones in a competition coordinated by the city council and decided by a public jury associated with any falla, all Fallas are set on fire, pardoning before the best figures of each one of them chosen by popular acclaim, as shown in Figure \ref{fig:figFallas2}.

It is a deeply rooted and very popular festival, which mobilizes tens of thousands of Valencians in all neighbourhoods and streets of the capital. The combination of art, various shows, parades, fireworks, music, gastronomic offerings, and long nights of festivities, make it a unique festival, which attracts countless tourists every year \citep{mestre2008image}. The popular version of the origin of Fallas according to the Marquis of Cruïlles, was initiated by the carpenters' guild that burned on the eve of the day of their patron Saint Joseph, in a purifying bonfire, the shavings and leftover junk, cleaning the workshops before the spring \citep{i2006estudios}. In addition, they burned their "parots" (structures from which hung the candles that gave them light) since with the end of winter and the arrival of spring, and as the days became longer, they were no longer needed. According to this theory, popular inventiveness gave human form to these parots.

The truth is that Fallas is a complete festival that mixes traditions such as the "mascletàs" - combined firing of thousands of firecrackers to cause a spectacular roar that makes buildings vibrate - or fire castles - which combine the sound of firecrackers with colour combinations caused by the combustion of gunpowder-, the realization of large artistic monuments such as Fallas themselves -some of them more than 30 meters high-, or religious activities such as the Offering to the Virgin of the Desamparados, to which are added the individual activities of each of the more than 350 fallas commissions in the city, with festivals, children's games, contests, and a long etcetera of activities. Fallas are above all an integral part of the Valencian idiosyncrasy and a cultural symbol rooted intergenerationally and celebrated by both religious and non-religious people. A festivity that vertebrates the local spirit of belonging to a community and that works as a powerful identity icon.

\section{The gunpowder, symbol of the Valencian festivity}

We could say that gunpowder is one of the essential elements of the Valencian festival, as it is present in the most deeply rooted festive traditions, as is the case of the celebration of Moros i Cristians and, of course, of Fallas. During these days, the smell of gunpowder is present from morning to night, as the falleros and falleras start at dawn exploding their firecrackers and so continue during noon, afternoon and even at night \citep{costa2002festive2}.

From the despertà, through the mascletà, the Nit del Foc, or the day of the Cremà itself. In each of these acts, the protagonist is the gunpowder, and without it, the festivities would probably be very different: they would not create the same spirit. And in each of these acts, it must be said that the gunpowder takes on a different personality, because if in the mascletà the smell of gunpowder shares the limelight with the noise, in the Nit del Foc it does so with colour and light. No one can resist the beauty of seeing the night sky illuminated with a mixture of colours and incredible shapes, and of course, with more alcohol than an overflowing river in a summer monsoon \citep{andres2016estimation}.

In each mascletà numerous kilograms of gunpowder are launched. The 'cordà' and the 'correfocs' are other traditional moments in which fire and gunpowder are the protagonists. These usually take place during the popular festivities of most of the Valencian towns. Of course, this is highly criticized by the environmental community for the production of smoke and waste, and by the community committed to animal ethics for the noise that frightens the animals.

\subsection{The drunk fireworks}
\begin{figure}
	\includegraphics[width=\columnwidth, trim={0 0 0 0.5cm},clip]{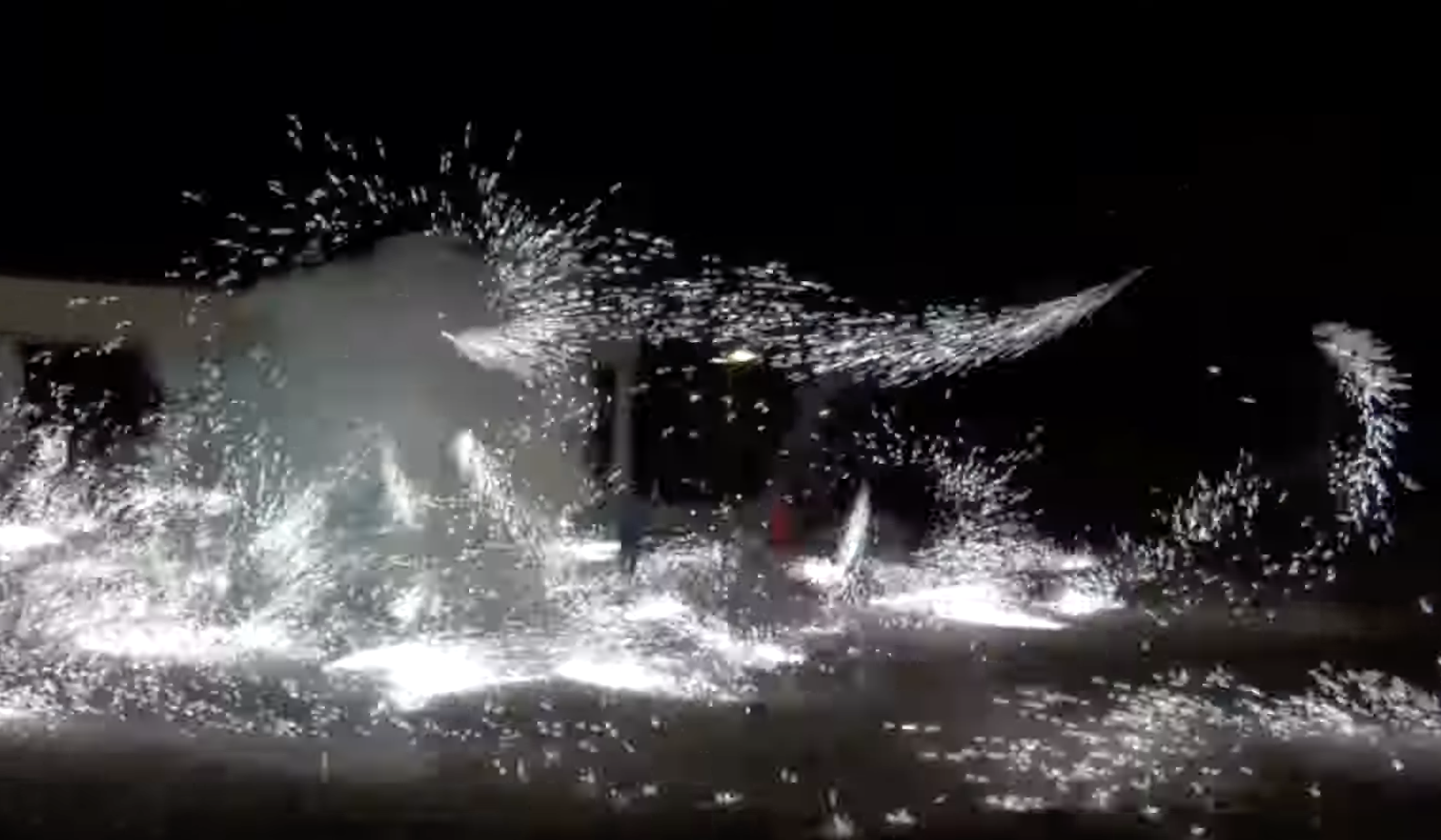}
	\centering
	\caption{Launching of multiple drunken fireworks. It is possible to appreciate the great speed of the same ones.}
	\label{fig:figDrunk}
\end{figure}

One of the most iconic and charismatic fireworks of this celebration are the well-known "borrachos" or "carretillas", in English crazy or drunk fireworks \citep{villarroya1992ciudad}. This type of dynamic firecracker consists of a hard cylinder that resists the combustion of gunpowder, resembling the propulsion of a rocket leaving sparks behind, but without flight control systems. This makes the motion of the drunk firework energetic and unpredictable, as evidenced in Figure \ref{fig:figDrunk}. This type of pyrotechnic device has been persecuted by the authorities because of its dangerousness, and there was even a European law proposal for its prohibition in 2007 \citep{martin2014analytical}.

\subsection{The popular myth: "If you run, it chases you"}

The knowledge acquired outside of school and apart from the academy, especially that one originated in the popular classes, is considered by the dominant discourse to be useless because it has no scientific basis. This discourse is preconceived and constitutes a silencer of the knowledge that the popular young and adult classes have built throughout their histories and life experiences \citep{crouch1998street}.

Ingrained as popular knowledge, there is the belief that drunk fireworks chase those who try to run away from them. This myth has been passed down from generation to generation and is widely used as advice not to get hurt by this kind of pyrotechnic device. Mothers and fathers, nowadays, still advise their children not to move if they see or hear a drunk firework. Some people even say that these small rockets smell fear, others explain that it is an effect produced by the air flow when running (also known as drafting or slipstreaming \citep{flynn2016effect}. Is it a widespread confirmation bias or a complex aerodynamic phenomenon? The following section details a numerical simulation of the simplest explanation without taking into account air forces: when running, a person passes through more volume and is, therefore, more exposed to being hit by a moving firework. However, the motion of these fireworks is chaotic, so apparently, at any given instant any section of the volume through which they move would have the same probability of being hit. What is the correct answer?

\section{Simulating the runaway}

\begin{figure*}
	\includegraphics[width=2\columnwidth]{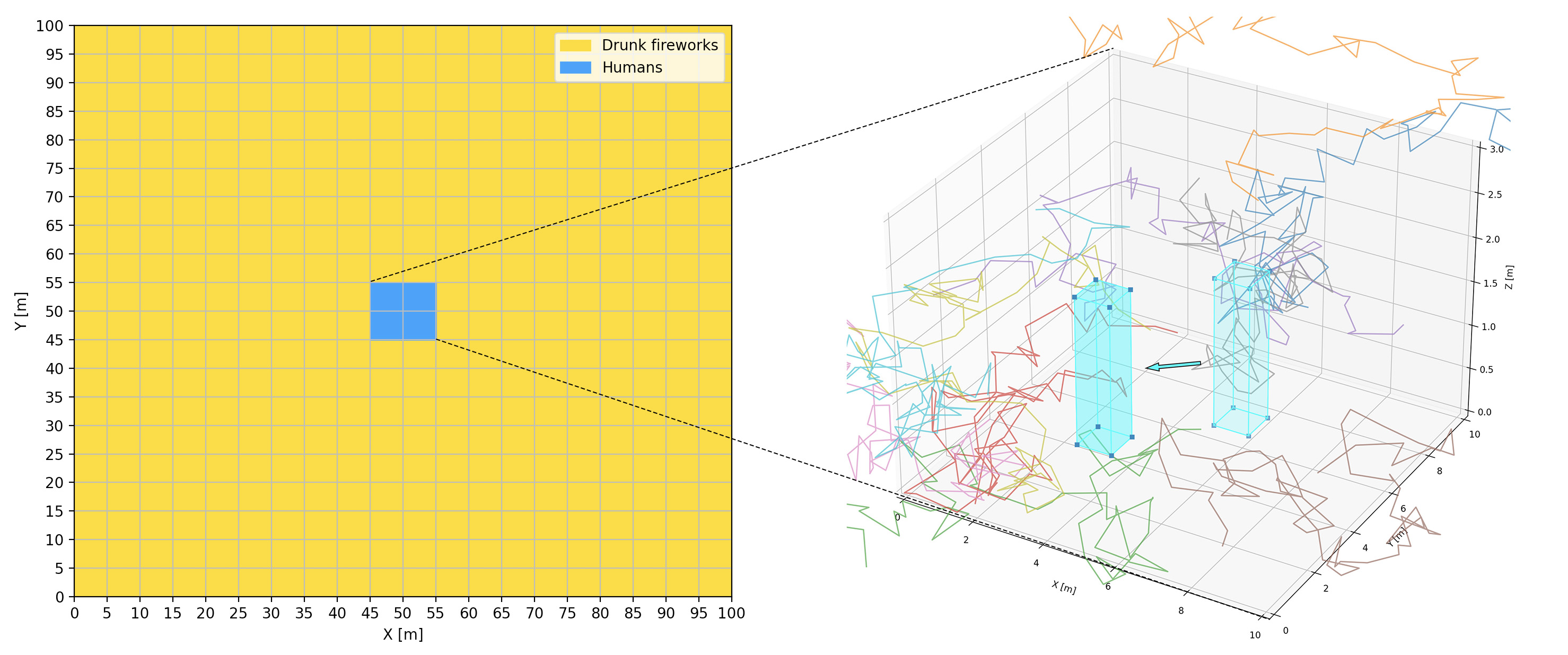}
	\centering
	\caption{Left: perpendicular view of the x-y plane with the representation of the areas where people and fireworks are randomly distributed. Right: A representation of linear and chaotic motion respectively.}
	\label{fig:figSim}
\end{figure*}

The simulations consist of an initial volume of 100x100x3 meters, being randomly distributed in any position the fireworks drunk. People are modelled by a parallelepiped of 1x1x1.7 meters initially distributed in the centre of the total volume at ground level on a square of 10 meters side. Two parallel simulations of 12 seconds each (which is the average duration of this type of fireworks) are performed: 1) keeping the people standing in their initial position; 2) imposing to the people a uniform rectilinear plane motion of 5 m/s. The fireworks follow a random walk at 4 times the human's speed with the only constraint that their height is not less than zero, i.e., that they do not cross the ground. Figure \ref{fig:figSim} shows the initial volume with human and firework distributions as well as the respective motions.

For each time iteration, it is calculated if any of the fireworks has a value of z<1.70, and if so, it is computed if its projection in the x-y plane falls within the base of the volume representing each person as impact condition.

In this sense, Monte Carlo simulations \citep{mooney1997monte} are applied to obtain 1000 calculations to correctly evaluate the statistical significance of impact probability with 10, 50, 100, 500 and 1000 initial drunk fireworks and a single person (with and without motion simultaneously).

\begin{figure}
	\includegraphics[width=\columnwidth]{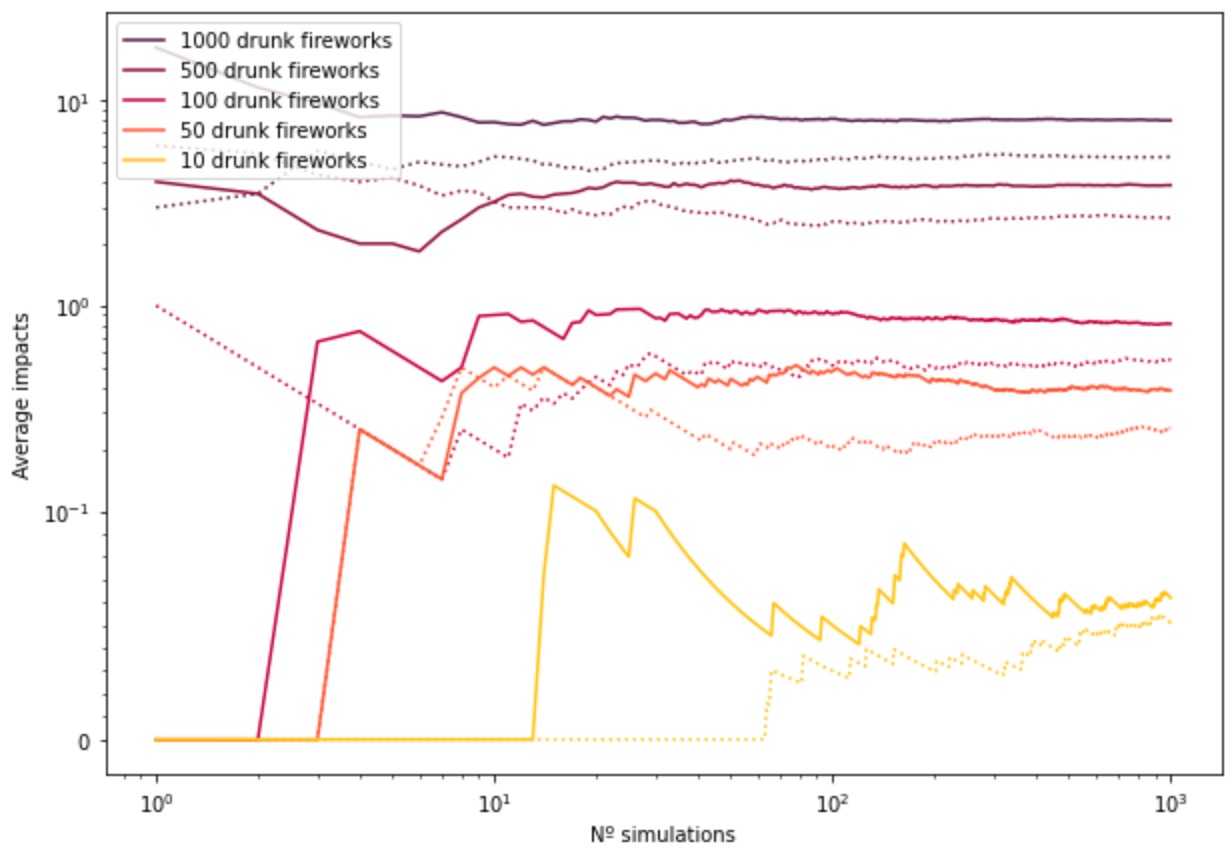}
	\centering
	\caption{Evolution of average impacts in the Monte Carlo simulations. The dotted lines represent people in static position.}
	\label{fig:figEvol}
\end{figure}

It can be observed in Figure \ref{fig:figEvol} the evolution of average impacts as a function of number of simulations. It is clear that the more fireworks there are, the greater the probability of being hit by one of them, and that regardless of the number of fireworks, a person who stays static will always receive less impact, increasing the benefit of not moving with the increase of pyrotechnic devices.

Table \ref{tab:results} shows the mean results with standard deviations (SD) for 1000 Monte Carlo simulations with 10, 50, 100, 500 and 1000 initial drunk fireworks for the cases of one person moving (M) and one person not moving (NM). 

\begin{table}
	\centering
	\caption{Average results with standard deviations (SD) for 1000 Monte Carlo simulations with 10, 50, 100, 500 and 1000 initial drunk fireworks (FW) for the cases of one person moving (M) and one person not moving (NM).}
	\label{tab:results}
	\begin{tabular}{lcccc} 
		\hline
		Nº FW & Impacts (M) & SD (M) & Impacts (NM) & SD (NM)\\
		\hline
		10 & 0.062 & 0.249 & 0.052 & 0.222\\
		50 & 0.386 & 0.640 & 0.251 & 0.508\\
		100 & 0.815 & 0.869 & 0.543 & 0.713\\
		500 & 3.851 & 1.869 & 2.673 & 1.611\\
		1000 & 7.965 & 2.881 & 5.280 & 2.391\\
		\hline
	\end{tabular}
\end{table}

\begin{figure}
	\includegraphics[width=\columnwidth]{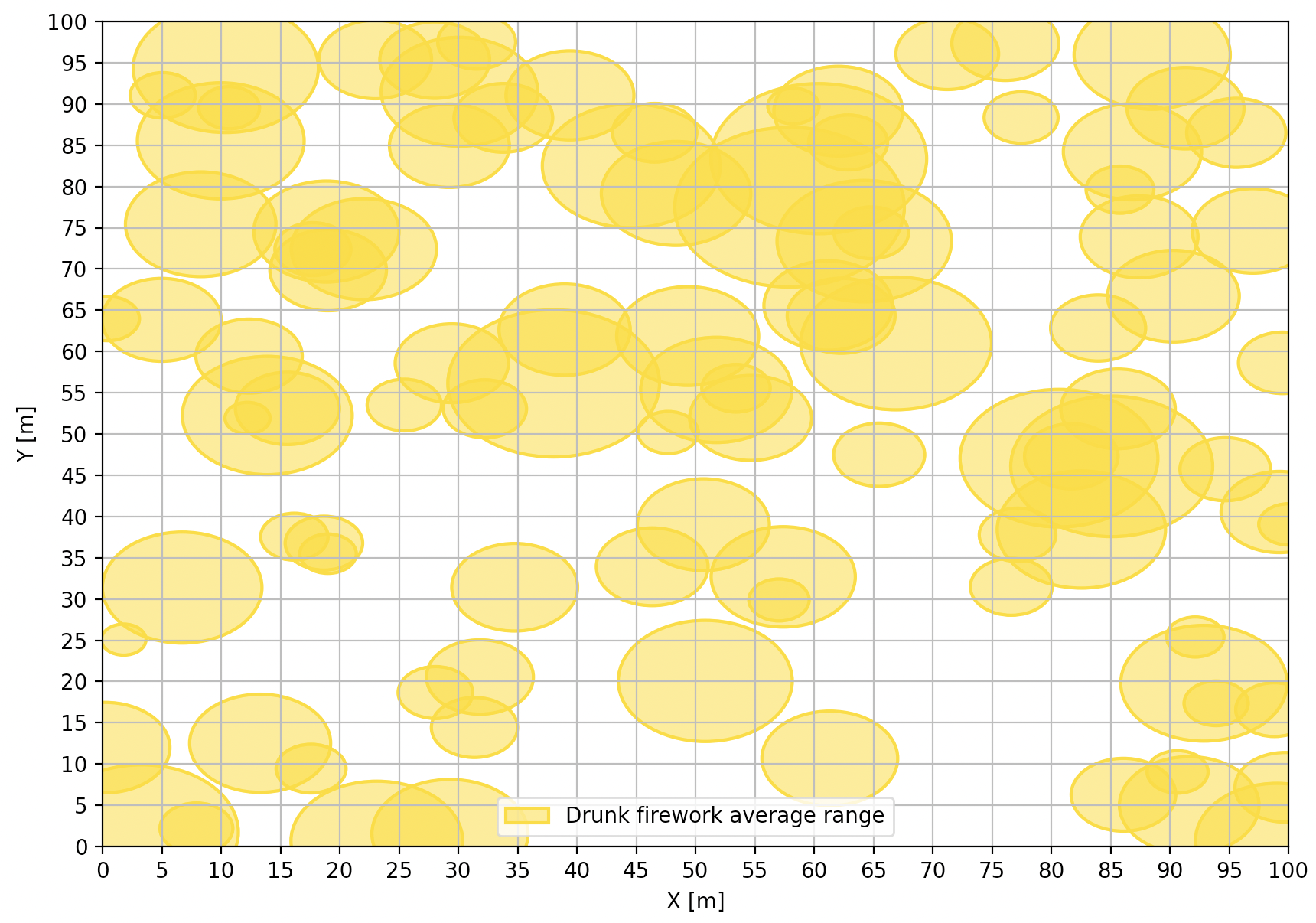}
	\centering
	\caption{Randomly distributed average motion ranges of drunk fireworks.}
	\label{fig:figRange}
\end{figure}

By distributing the drunk fireworks randomly and having a chaotic motion for a short time, they will move more or less around their initial position, thus defining clusters of average ranges of motion, as can be seen in Figure \ref{fig:figRange}. Except for the cases with very large numbers of fireworks, it is more likely that a person randomly positioned in the centre of the studied volume is not within the range of any drunk firework or at least only exposed to one or a few impact ranges, so when moving it increases the probability of intersecting more sections with risk of impact.

\section{Discussion and Conclusions}

It is important to recognize that popular knowledge arises from very diverse life experiences and ways of knowing the world that are produced outside the formal spaces of education, i.e., that are inherited or have their origin in popular milieus, in social movements and/or in religious, ethnic, associative spheres, modern social tribes... It is about knowledge that could contribute to the development of all the potentials and dimensions of the human being. Although they are produced from the individual experience of each person, added to shared issues of the commons they give shape to different identities (in this case myths associated with belonging to an idiosyncrasy and a local festivity), and may also be functional and intended to protect loved ones (e.g., to prevent people from getting hurt by uncontrolled fireworks).

The experience transmitted from generation to generation by accumulation of individual experiences is an interesting source of possible knowledge to be analyzed, which may contain valuable information without the necessary intervention of the scientific method. These urban legends and hearsay tales often serve a function beyond their truthfulness since they directly appeal to human emotions and on many occasions accurately describe a complex reality. Obviously, some myths are traditional burdens that try to maintain a conservative social inertia, often falling into unjustified prejudices and sustaining oppressive dynamics.

In the specific case of the Valencian Fallas festivities and the myth about the drunk fireworks analyzed in this work, popular wisdom seems to be correct in some way. Without a computational fluid dynamics study we can not confirm if a drunk firework really chases a person running away by drafting or slipstreaming effects, but at the very least, only by mere probability of intersection of volumes, if you run near a drunk firework, you are more likely to be hit.

\section*{Acknowledgements}

I thank my very beloved life partner Elisa for showing me the Valencian festivities of Fallas despite her ideological reticence. Thanks also to Ana María for stimulating the debate that has motivated this work. Thanks to Manu for his affectionate and ever-present company.


\section*{Data Availability}

The data underlying this article will be shared on reasonable request to the corresponding author



\bibliographystyle{mnras}
\bibliography{example} 





\bsp	
\label{lastpage}
\end{document}